\documentclass[twocolumn,superscriptaddress,floats,prd,nofootinbib,showpacs]{revtex4-2}
\usepackage{anyfontsize} 
\usepackage{stmaryrd}
\usepackage{mathrsfs}
\usepackage{float}
\usepackage{setspace}
\usepackage{amsmath,amssymb,amsfonts}
\usepackage{graphicx}
\usepackage{subfigure}
\usepackage{color} 
\usepackage{fancyhdr}
\usepackage{hyperref} 
\usepackage{enumerate}
\usepackage{CJKutf8}
\usepackage{color}
\usepackage{multirow,booktabs}
\usepackage{enumitem}
\newcommand{\Rmnum}[1]{\uppercase\expandafter{\romannumeral #1}}  
\newcommand{\bea}{\begin{eqnarray}}
	\newcommand{\eea}{\end{eqnarray}}

\begin{document}
\begin{CJK}{UTF8}{gbsn}

	\title{LQG-inspired black hole solutions in (2+1) dimensions}
	\author{Zijian Shi}
	\affiliation{School of Physics and Optoelectronics, South China University of Technology, Guangzhou 510641, China}
	\author{Xiangdong Zhang}\email{Corresponding author. scxdzhang@scut.edu.cn}
	\affiliation{School of Physics and Optoelectronics, South China University of Technology, Guangzhou 510641, China}
	\begin{abstract}
		It is well known that in (2+1)-dimensional general relativity, black hole solutions exist only in the anti-de Sitter case—the so-called Banados-Teitelboim-Zanelli (BTZ) black hole. In this letter, we construct, for the first time, three independent LQG-inspired frameworks for (2+1)-dimensional black holes: the $\bar{\mu}$-scheme holonomy corrections, the quantum Oppenheimer-Snyder collapse, and a covariant effective model. Remarkably, all three approaches yield the same qualitative picture: when LQG-inspired corrections are included, we find three distinct types of solutions corresponding to asymptotically flat, anti-de Sitter, and de Sitter geometries. Our results demonstrate that the asymmetry between classical (2+1)-dimensional and (3+1)-dimensional solutions is eliminated upon incorporating effective quantum gravity effects.
		
	\end{abstract}

	\maketitle


	{\it Introduction---}Research in three-dimensional gravity has been highly active over recent decades, primarily because it offers a simplified framework that nevertheless retains essential conceptual features of four-dimensional general relativity \cite{Carlip:2004ba}. 
	It is well established that pure three-dimensional Einstein gravity in vacuum has no local degrees of freedom. Thus, the discovery by Banados, Teitelboim, and Zanelli in 1992 that black hole solutions can exist in (2+1)-dimensional anti-de Sitter spacetime was groundbreaking \cite{Banados:1992wn}. 
	The resulting BTZ black hole indeed exhibits hallmark properties of black holes \cite{Carlip:1995qv}: 
	it possesses an event horizon, can arise from gravitational collapse \cite{Ross:1992ba}, 
	and displays thermodynamic behavior analogous to that of (3+1)-dimensional black holes \cite{Banados:1992wn, Kubiznak:2016qmn}, including entropy and Hawking temperature. The BTZ black hole has significantly advanced our understanding of both classical and quantum aspects of gravity \cite{Carlip:1995qv,Banados:1992gq, Carlip:2005zn}, particularly in contexts where a full quantum theory of gravity in 3+1 dimensions remains out of reach. However, a key distinction lies in the absence of asymptotically flat or de Sitter solutions in (2+1) dimensions, which results in a notable asymmetry between three-dimensional and four-dimensional black hole physics.
	
	On the other hand, general relativity (GR) faces fundamental limitations when spacetime curvature approaches the Planck scale, particularly in the vicinity of spacetime singularities. Such a breakdown calls for a quantum gravity framework capable of unifying quantum mechanics with GR to resolve these singularities. Loop quantum gravity (LQG), characterized by its background-independent formulation, stands out as a prominent candidate for quantum gravity \cite{Ashtekar:2004eh, Rovelli:2004tv, Thiemann:2007pyv, Han:2005km}. In particular, the intrinsic discreteness of LQG has led to remarkable progress, including the resolution of cosmological big-bang and black hole singularities \cite{Bojowald:2001xe, Ashtekar:2005qt, Boehmer:2007ket, Chiou:2008nm}. It is thus tempting to employ such an ideal framework to investigate issues related to three-dimensional quantum black holes. Actually, 2+1-dimensional LQG has been well-established since its original formulation by Ashtekar \textit{et al.} in 1989 \cite{Ashtekar:1989qd}. 
	A wide range of its properties has been explored, some of which even offer valuable insights into the corresponding problems in (3+1) dimensions. \cite{Thiemann:1997ru, Ma:2001dm, Freidel:2002hx, Noui:2004iy, Bonzom:2008tq, Meusburger:2008bs, Perez:2010pm, Bonzom:2011hm, Frodden:2012nu, Zhang:2014xqa}. 
	
	Currently, the application of loop quantization techniques to black hole models is not uniquely determined, even for the simplest Schwarzschild black hole. Substantial progress has been achieved in recent years, both in effective Hamiltonian constructions within canonical approach (including the $\bar{\mu}$ scheme) and in quantum collapsing models. Loop quantum gravity can be formulated as a Hamiltonian theory via canonical quantization. This approach relies on an n+1 decomposition of spacetime, in which the manifold is foliated into spatial hypersurfaces parameterized by a time coordinate. This structure obscures general covariance. LQG corrections typically involve replacing polynomial curvature terms in the classical Hamiltonian with bounded functions \cite{Bojowald:2015zha}. 
	As emphasized in \cite{Belfaqih:2024vfk}, this procedure is usually implemented by hand, rather than being rigorously derived from first principles, and the covariance of the resulting models is seldom verified or even acknowledged as a relevant issue. Accordingly, the problem of covariance has remained a long-standing open question in LQG  \cite{Bojowald:2015zha, Belfaqih:2024vfk, Bojowald:2011aa, Tibrewala:2013kba, Wu:2018mhg, Bojowald:2019dry, Bojowald:2020unm, Han:2022rsx, Gambini:2022dec, Bojowald:2022zog, Alonso-Bardaji:2021yls, Alonso-Bardaji:2022ear, Ashtekar:2023cod, Giesel:2023hys, Bojowald:2024beb, Bojowald:2023djr, Bojowald:2023vvo, Alonso-Bardaji:2023vtl, Bojowald:2024ium}. 
	Several strategies have been developed to tackle this problem \cite{Yang:2025ufs}. One such approach consists in expressing the relevant equations using fully tensorial notation, as exemplified by the quantum Oppenheimer-Snyder model \cite{Lewandowski:2022zce, Shi:2024vki}. A prominent realization is the construction of an effective Hamiltonian constraint that manifestly preserves covariance \cite{Alonso-Bardaji:2023vtl, Zhang:2024khj, Zhang:2024ney, Belfaqih:2024vfk}. 
	In particular, the systematic method proposed in \cite{Zhang:2024khj, Zhang:2024ney} translates this covariance requirement into a set of equations whose solutions directly yield such covariant constraints. General solutions to the covariance equations, as well as extensions to non-vacuum scenarios, have also been explored \cite{Yang:2025ufs, Zhang:2025ccx}. One may also introduce a suitable matter field to fix the diffeomorphism gauge \cite{Han:2022rsx, Giesel:2023tsj}. Moreover, this idea is also used to obtain
a regular black hole from an effective Lagrangian for quantum Einstein gravity \cite{Bonanno24}.

To date, all such investigations have been carried out exclusively in 3+1 dimensions. In this letter, we aim to investigate black hole solutions in (2+1)-dimensional loop quantum gravity for the first time, using three independent LQG-inspired approaches: the $\bar{\mu}$-scheme holonomy corrections, the quantum Oppenheimer-Snyder collapse, and a covariant effective model. These approaches provide complementary perspectives and collectively establish the robustness of the results. For readability, detailed calculations of the $\bar{\mu}$ scheme and the quantum Oppenheimer-Snyder model are provided in the Supplemental Material.

    {\it Hamiltonian formulation---}Our starting point is the action \cite{Carlip:1998uc}
    \begin{equation}
		\mathcal{S} = \frac{1}{2\kappa} \int d^3X \sqrt{-\det(g)} (R - 2\Lambda) ,
	\end{equation}
	where $\kappa = 8 \pi G$, with $G$ Newton's constant and $\Lambda$ the cosmological constant.

	We consider the circularly symmetric situation. 
	The spacetime is defined on a three-dimensional manifold $\mathcal{M}$ with 
	the topological structure $\mathcal{M} \cong \mathbb{R} \times \Sigma$. 
	The spatial section decomposes as $\Sigma \cong \mathcal{M}_1 \times \mathbb{S}^1$, 
	with $\mathcal{M}_1$ being a one-dimensional manifold (\textit{e.g.}, $\mathbb{R}^+$) and $\mathbb{S}^1$ the 1-sphere. 
	Coordinates on $\mathcal{M}$ are chosen as $(t, x, \phi)$. 
	The corresponding phase space is spanned by the canonical pairs $(K_1, E^1)$ and $(K_2, E^2)$, 
	which obey nontrivial Poisson brackets dictated by the symplectic structure: 
	$\{K_1(x), E^1(y)\} = 4G\delta(x, y)$ and $\{K_2(x), E^2(y)\} = 4G\delta(x, y)$.
	It is worth noting that $K_1$ and $E^2$ are scalar densities of weight one, 
	while $K_2$ and $E^1$ are scalars. Then classically a metric $g_{\rho \sigma}$ can be constructed on $\mathcal{M}$ as 
	\begin{equation}\label{metric}
		\mathrm{d}s^2 = -N^2 \mathrm{d}t^2 + (E^2)^2 (\mathrm{d}x + N^x \mathrm{d}t)^2 + (E^1)^2 \mathrm{d}\phi^2.
	\end{equation}

	The dynamics are determined by the Hamiltonian constraint $H$ and the diffeomorphism constraint $H_x$. 
	Classically, they take the following form (see Supplemental Material for detailed derivations) \cite{Wald:1984rg, Liang:2023ahd, Thiemann:2007pyv, Kelly:2020uwj}: 
	\begin{equation}\label{H}
		H = -\frac{K_1 K_2}{4G} 
		- \frac{(\partial_x E^1) \partial_x E^2}{4G(E^2)^2} 
		+ \frac{\partial_x^2 E^1}{4GE^2} 
		+ \frac{E^1 E^2 \Lambda}{4G},
	\end{equation}
	\begin{equation}\label{Hx}
		H_x = \frac{1}{4G}\left(E^2 \partial_x K_2 - K_1 \partial_x E^1\right).
	\end{equation}
	The constraint algebra reads
	\begin{subequations}\label{CA}
		\begin{align}
			\{H_x[N_1^x], H_x[N_2^x]\} &= H_x[N_1^x \partial_x N_2^x - N_2^x \partial_x N_1^x], \label{CA1} \\
			\{H[N], H_x[N^x]\} &= -H[N^x \partial_x N], \label{CA2} \\
			\{H[N_1], H[N_2]\} &= H_x[S(N_1 \partial_x N_2 - N_2 \partial_x N_1)], \label{CA3}
		\end{align}
	\end{subequations}
	with $S = (E^2)^{-2}$ being the structure function. 
	We have adopted the notation $F[g] \equiv \int F(x)g(x)\mathrm{d}x$ for the smeared constraints. 
	The Hamiltonian equations of motion (EOM) are given by $\dot{\mathscr{F}} = \{\mathscr{F}, H[N] + H_x[N^x]\}$, 
	where we use $\mathscr{F}$ to denote canonical variables $K_I$ and $E^I$. 
	After choosing a lapse function $N$ and a shift vector $N^x \partial_x$, 
	solving the EOM along with the constraint equations $H=0$ and $H_x=0$ 
	yields fields $K_I(t, x)$ and $E^I(t, x)$ on $\mathcal{M}$.

    {\it Effective dynamics with holonomy corrections---} Before addressing the general covariance issue, it is instructive to consider a concrete realization of LQG-inspired corrections within a reduced phase space. Following the standard polymerization prescription in loop quantum cosmology (LQC), we partially fix the gauge by setting $E^1 = x$. The diffeomorphism constraint $H_x$ then enforces $K_1 = E^2 \partial_x K_2$, and dynamical preservation of the gauge fixes the shift vector as $N^x = -N K_2$. The reduced Hamiltonian constraint reads 
    \begin{equation}
    H = -\frac{K_2 E^2 \partial_x K_2}{4G} - \frac{\partial_x E^2}{4G(E^2)^2} + \frac{x E^2 \Lambda}{4G}.
    \end{equation}
    Following the $\bar{\mu}$ scheme, the polymerization of the connection component $K_2$ is implemented by the replacement
    \begin{equation}
    K_2 \;\longrightarrow\; \frac{x}{\zeta}\sin\left(\frac{\zeta K_2}{x}\right),
    \end{equation}
    where $\zeta = \gamma L$ is a quantum parameter. Here, $\gamma$ is the Barbero-Immirzi parameter and 
	$\begin{aligned}L=4\sqrt{3}\pi\ell_{\mathrm{p}}\end{aligned}$ 
	is the minimal length gap, with 
	$\ell_{\mathrm{p}}$ being the Planck length \cite{Zhang:2014xqa}. This yields the effective Hamiltonian constraint
    \begin{equation}
    H^{(LQG)} = -\frac{1}{4G}\left[\frac{E^2}{2}\partial_x\left(\frac{x^2}{\zeta^2}\sin^2\frac{\zeta K_2}{x}\right) + \frac{\partial_x E^2}{(E^2)^2} - x E^2 \Lambda\right].
    \end{equation}
    Solving the resulting equations of motion together with $H^{(LQG)}=0$ in the Painlev\'e-Gullstrand-like gauge $N=1$ gives
    \begin{equation}
    \mathrm{d}s^2 = -f(x)\,\mathrm{d}T^2 + 2\sqrt{1-f(x)}\,\mathrm{d}T\,\mathrm{d}x + \mathrm{d}x^2 + x^2\,\mathrm{d}\phi^2,
    \end{equation}
    with
    \begin{equation}\label{barmufx}
    f(x) = -8Gm - \Lambda x^{2} + \frac{\zeta^2}{x^{2}} (1+8Gm + \Lambda x^{2})^2.
    \end{equation}
    To keep the notation uniform throughout the paper, we denote the time coordinate by $T$ in the Painlev\'e-Gullstrand-like gauge. For $\Lambda=0$, the solution is asymptotically flat and admits a cosmological horizon. The detailed derivation is provided in the Supplemental Material.

    {\it Quantum Oppenheimer-Snyder collapse---} An independent and dynamically distinct approach to effective quantum black holes is provided by the quantum Oppenheimer-Snyder model \cite{Lewandowski:2022zce, Shi:2024vki}. Consider a disk of pressureless dust collapsing in (2+1) dimensions. The interior spacetime is described by a spatially flat Friedmann-Robertson-Walker (FRW) metric, whose dynamics is governed by the LQC-modified Friedmann equation \cite{Zhang:2014xqa}
    \begin{equation}
    \mathcal{H}^2 = \kappa\left(\rho + \frac{\Lambda}{\kappa}\right)\left(1 - \frac{\rho + \Lambda/\kappa}{\rho_c}\right),
    \end{equation}
    where $\mathcal{H} = \dot{a}/a$ is the Hubble parameter, $\rho$ is the dust energy density, and $\rho_c = 1/(\kappa\zeta^2)$ is the critical density inherited from LQC. Matching the interior to an exterior static vacuum spacetime via the Darmois-Israel junction conditions yields the exterior metric
    \begin{equation}
    \mathrm{d}s^2 = -f(x)\,\mathrm{d}t^2 + f(x)^{-1}\,\mathrm{d}x^2 + x^2\,\mathrm{d}\phi^2,
    \end{equation}
    with
    \begin{equation}\label{qOSfx}
    f(x) = -8Gm - \Lambda x^{2} + \frac{\zeta^2}{x^{2}}(1+8Gm+\Lambda x^{2})^2.
    \end{equation}
    Remarkably, this $f(x)$ is of the same form as that obtained in the $\bar{\mu}$ scheme above. We note that for $\Lambda=0$ the same phenomenon—a cosmological horizon—again emerges. The full derivation is provided in the Supplemental Material.

    {\it Covariance issue---}The above two approaches provide concrete realizations of LQG effects. However, neither of them addresses covariance systematically within the Hamiltonian framework. To fill this gap, we now restart from the full Hamiltonian formulation presented at the beginning to analyze covariance from first principles.
    
    The evolution of a point in phase space and 
	evolution of the corresponding fields in spacetime can be connected, 
	allowing the EOM to be cast in the form of a Lie derivative: $\mathcal{L}_{\mathfrak{N}} \mathscr{F} = \{\mathscr{F}, H[N]\}$, 
	where $\mathfrak{N} = \partial_t - N^x \partial_x$. 
	Then, we consider an infinitesimal gauge transformation generated by 
	$H[\alpha N] + H_x[\beta^x]$, 
	where $\alpha$ is an arbitrary smeared function and 
	$\beta^x \partial_x$ is an arbitrary smeared vector field. 
	After a calculation similar to \cite{Zhang:2024ney}, it can be checked that: 
	\begin{equation}\label{covar}
		\delta g_{\rho \sigma} \mathrm{d}x^{\rho} \mathrm{d}x^{\sigma} = \mathcal{L}_{\alpha \mathfrak{N} + \beta} (g_{\rho \sigma} \mathrm{d}x^{\rho} \mathrm{d}x^{\sigma}).
	\end{equation}
	It means that the infinitesimal gauge transformation for $g_{\rho \sigma}$ 
	generated by $H[\alpha N] + H_x[\beta^x]$ in phase space can be identified as 
	the diffeomorphism transformation generated by the vector field 
	$\alpha \mathfrak{N} + \beta$ in spacetime. 
	This result establishes the covariance of the classical theory with respect to the metric $g_{\rho \sigma}$. 

	We now consider the quantum case. 
	Following a common approach in LQG \cite{Zhang:2024khj, Belfaqih:2024vfk}, we preserve the classical form of the diffeomorphism constraint, 
	which continues to generate spatial diffeomorphism transformations. 
	However, the Hamiltonian constraint is modified into an effective version, $H_{\mathrm{eff}}$, 
	to incorporate quantum gravity effects, and its specific form is to be determined. 
	The constraints are required to be first-class, 
	and the constraint algebra is assumed to deviate from the classical one only by 
	a $\mu$-dependent correction that accounts for quantum influences \cite{Bojowald:2011aa, Giesel:2023tsj, AlonsoBardaji:2023bww, Zhang:2024khj}: 
	\begin{subequations}\label{qCA}
		\begin{align}
			\{H_x[N_1^x], H_x[N_2^x]\} &= H_x[N_1^x \partial_x N_2^x - N_2^x \partial_x N_1^x], \label{qCA1} \\
			\{H_{\mathrm{eff}}[N], H_x[N^x]\} &= -H_{\mathrm{eff}}[N^x \partial_x N], \label{qCA2} \\
			\{H_{\mathrm{eff}}[N_1], H_{\mathrm{eff}}[N_2]\} &= H_x[\mu S(N_1 \partial_x N_2 - N_2 \partial_x N_1)], \label{qCA3}
		\end{align}
	\end{subequations}
	where structure function becomes $\mu S$ and $\mu$ can be a function of $K_I$ and $E^I$. 
	The classical structure function is interpreted as 
	the $(x, x)$-component of the inverse spatial metric, 
	so the constraint algebra encodes the hypersurface-deformation properties. 
	For the algebra \eqref{qCA} to retain this geometric meaning, 
	one must define an effective metric $g_{\rho\sigma}^{(\mu)}$ that generalizes the classical one \cite{Bojowald:2011aa, Zhang:2024khj, Zhang:2024ney}: 
	\begin{equation}\label{qmetric}
		\mathrm{d}s^2 = -N^2 \mathrm{d}t^2 + \frac{(E^2)^2}{\mu} (\mathrm{d}x + N^x \mathrm{d}t)^2 + (E^1)^2 \mathrm{d}\phi^2.
	\end{equation}
	Then, we can compute the infinitesimal gauge transformation for $g_{\rho\sigma}^{(\mu)}$ generated by 
	$H[\alpha N] + H_x[\beta^x]$ in analogy with the classical procedure. 
	In order to derive the covariant result: 
	\begin{equation}
		\delta g_{\rho \sigma}^{(\mu)} \mathrm{d}x^{\rho} \mathrm{d}x^{\sigma} = \mathcal{L}_{\alpha \mathfrak{N} + \beta} (g_{\rho \sigma}^{(\mu)} \mathrm{d}x^{\rho} \mathrm{d}x^{\sigma}),
	\end{equation}
	we establish the necessary and sufficient conditions: 
	\begin{enumerate}[label=(\roman*)]
		\item $H_{\mathrm{eff}}$ is independent of $\partial_x^n K_1$ for all $n \geq 1$;
		\item the following condition holds for any phase-space-independent functions $\alpha$ and $N$: 
		\begin{equation*}
			\alpha\{\mu S, H_{\mathrm{eff}}[N]\} = \{\mu S, H_{\mathrm{eff}}[\alpha N]\}.
		\end{equation*}
	\end{enumerate}
	Detailed calculations are available in the Supplemental Material. 

	Since $H_{\mathrm{eff}}$ is a scalar density of weight one, it allows us to 
	make the ansatz $H_{\mathrm{eff}}=E^2 F$, 
	where $E^2$ is also a scalar density of weight one and $F$ is a scalar. 
	Consequently, $F$ is restricted to being a function of the 
	basic scalars derived from $K_I$, $E^I$, and their derivatives. 
	The discussion on basic scalars in \cite{Zhang:2024khj} also applies to our model. 
	Based on the covariance condition (i), 
	we exclude the derivatives of both $K_1$ and $K_2$, as they play similar roles in the model. 
	Additionally, the constraint algebra \eqref{qCA3} restricts derivatives of $E^I$ to the second order at most. 
	As a result, we screen out the following basic scalars in 2+1 dimensions: 
	\begin{equation}
		\begin{aligned}
			s_{1} &= E^{1}, \quad & s_{2} &= K_{2}, \quad & s_{3} &= \frac{K_{1}}{E^{2}}, \\
			s_{4} &= \frac{\partial_x E^{1}}{E^{2}}, \quad & s_{5} &= \frac{1}{E^{2}} \partial_x \left( \frac{\partial_x E^{1}}{E^{2}} \right).
		\end{aligned}
	\end{equation}
	Substituting the ansatz for $H_{\mathrm{eff}}$ into the constraint algebra \eqref{qCA3} and 
	imposing the necessary and sufficient conditions for covariance, 
	we obtain that $H_{\mathrm{eff}}$ takes the following form: 
	\begin{equation}\label{Heff}
		H_{\mathrm{eff}} = -E^2 \left[ \partial_{s_1} M_{\mathrm{eff}} + (\partial_{s_2} M_{\mathrm{eff}}) s_3 + \frac{\partial_{s_4} M_{\mathrm{eff}}}{s_4} s_5 + \mathcal{R} \right].
	\end{equation}
	Here, $\mathcal{R}(s_1, M_{\mathrm{eff}})$ is an arbitrary function, 
	where the cosmological constant term can be included, 
	and $M_{\mathrm{eff}}(s_1, s_2, s_4)$ is any solution to the following covariance equations: 
	\begin{equation}\label{CE1}
		\frac{\mu s_4}{16 G^2} = (\partial_{s_2} M_{\mathrm{eff}}) \partial_{s_2} \partial_{s_4} M_{\mathrm{eff}} - (\partial_{s_4} M_{\mathrm{eff}}) \partial_{s_2}^2 M_{\mathrm{eff}},
	\end{equation}
	\begin{equation}\label{CE2}
		(\partial_{s_2} \mu) \partial_{s_4} M_{\mathrm{eff}} - (\partial_{s_4} \mu) \partial_{s_2} M_{\mathrm{eff}} = 0.
	\end{equation}
	Eq. \eqref{CE2} implies that $\mu$ only depends on $s_1$ and $M_{\mathrm{eff}}$.

	{\it Covariant solutions---}As expected, the classical Hamiltonian constraint \eqref{H} can be recovered by setting $\mu=1$ in 
	the covariance equations to obtain: 
	\begin{equation}
		M_{\mathrm{cl}} = \frac{1}{2G} \left[ 1 + \frac{(s_2)^2}{4} - \frac{(s_4)^2}{4} \right],
	\end{equation}
	and simultaneously choosing the free function in Eq. \eqref{Heff} to be $\mathcal{R_{\mathrm{cl}}} = -\Lambda s_1 / 4G$. 
	
	To construct a viable framework for LQG-inspired black holes, 
	one can adopt the standard treatment of LQC, 
	which requires holonomy corrections \cite{Ashtekar:2006wn}. 
	The substitution of connections by their holonomies, 
	namely the polymerization procedure, 
	guarantees the background independence of the resulting theory \cite{Ashtekar:1995zh}. 
	In this covariant framework, the polymerization is implemented by the replacement $s_2 = K_2 \to \zeta^{-1} s_1 \sin\left(\frac{\zeta s_2}{s_1}\right)$ \cite{Zhang:2014xqa, Zhang:2023yps}. 
	With holonomy corrections incorporated, 
	we find the following solution to the covariance equations: 
	\begin{equation}\label{Meff}
		M_{\mathrm{eff}} = \frac{1}{2 G} + \frac{(s_1)^2 \sin^2\left(\frac{\zeta s_2}{s_1}\right)}{8G\zeta^2} - \frac{(s_4)^2}{8 G} e^{\frac{2i\zeta s_2}{s_1}}.
	\end{equation}
    Although $M_{\mathrm{eff}}$ and $H_{\mathrm{eff}}$ appear complex for non-zero $s_2=K_2$, in the static solution presented below, $K_2$ takes a complex value determined by solving the dynamics, and it is precisely this complex value that guarantees the reality of $M_{\mathrm{eff}}$ and the resulting effective metric.
	Ref. \cite{Yang:2025ufs} has considered a general scenario, 
	where the cosmological constant term in Eq. \eqref{Heff} is also subject to quantum correction, 
	and draws interesting conclusions. 
	However, in our model, substituting Eq. \eqref{Meff} into Eq. \eqref{CE1} yields $\mu = 1$, 
	which implies that the cosmological constant term can be only coupled to a constant. 
	Hence, without loss of generality, we only need to consider $\mathcal{R} = \mathcal{R_{\mathrm{cl}}}$. 

	To obtain the effective metric, we solve the dynamics in the stationary case \cite{Cafaro:2024vrw, Yang:2025ufs}. 
	We set $E^1 = x$ to fix the gauge from the diffeomorphism constraint. 
	Then, through the stationarity condition, $\{E^I, H_{\mathrm{eff}}[N] + H_x[N^x]\} = 0$, 
	the lapse function $N$ and shift vector $N^x \partial_x$ can be fixed. 
	Finally, we need to determine $E^2$, 
	which is associated with the constraint equations $H_{\mathrm{eff}} = 0$ and $H_x = 0$. 
	Different gauge choices result in variations in the final form of the metric. 
	In the gauge $N^x = 0$, 
	the effective metric \eqref{qmetric} becomes: 
	\begin{equation}\label{SWCgauge}
		\mathrm{d}s_{\mathrm{eff}}^2 = -f_{\mathrm{eff}}(x) \, \mathrm{d}t^2 + f_{\mathrm{eff}}(x)^{-1} \, \mathrm{d}x^2 + x^2 \, \mathrm{d}\phi^2,
	\end{equation}
	where 
	\bea\label{feff}
		f_{\mathrm{eff}}(x) = (-8Gm - \Lambda x^{2}) \left[1 + \frac{\zeta^{2}}{x^{2}} (-8Gm - \Lambda x^{2})\right],
	\eea
	with $m$ being an integration constant. Interestingly, when $\Lambda=0$, we obtain the asymptotically flat solution as
    \bea\label{vaccum}
    \mathrm{d}s_{\mathrm{eff}}^2 &=& -\left(-8Gm+\frac{64G^2m^2\zeta^{2}}{x^{2}}\right) \mathrm{d}t^2 \nonumber\\
    &+& \left(-8Gm+\frac{64G^2m^2\zeta^{2}}{x^{2}}\right)^{-1} \mathrm{d}x^2 + x^2\mathrm{d}\phi^2, \nonumber\\
    \eea
    which is clearly absent in the classical case. Moreover, $\Lambda$ can also adopt positive values, which also has no counterpart in the classical situation. Hence, the asymmetry between classical (2+1)-dimensional and (3+1)-dimensional solutions is eliminated upon incorporating effective quantum gravity effects.
	
When the Painlev{\'e}-Gullstrand-like \cite{Painleve, Gullstrand:1922tfa, Liu:2005hj} gauge $E^2=1$ is chosen, 
	the effective metric \eqref{qmetric} takes the form: 
	\begin{equation}\label{PGgauge}
		\mathrm{d}s_{\mathrm{eff}}^{2} = -\mathrm{d}T^{2} + \left[ \mathrm{d}x \pm \sqrt{1 - f_{\mathrm{eff}}(x)}  \mathrm{d}T \right]^{2} + x^{2} \mathrm{d}\phi^{2} .
	\end{equation}
	Solutions \eqref{SWCgauge} and \eqref{PGgauge} are equivalent under the coordinate transformation: 
	\begin{equation}
		\mathrm{d}t = \mathrm{d}T \mp \frac{\sqrt{1 - f_{\mathrm{eff}}(x)}}{f_{\mathrm{eff}}(x)} \mathrm{d}x.
	\end{equation}
	This equivalence underscores the inherent covariance of the theory.

\begin{figure}[!htb]
		\includegraphics [width=0.25\textwidth]{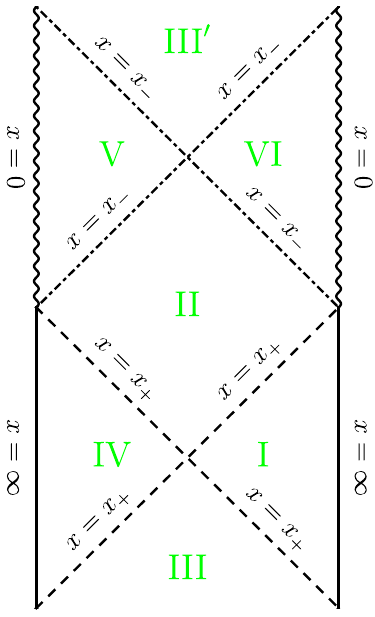}
		\caption{The Penrose diagram for $\Lambda<0$}
		\label{qBTZ1}
	\end{figure}

\begin{figure}[!htb]
		\includegraphics [width=0.25\textwidth]{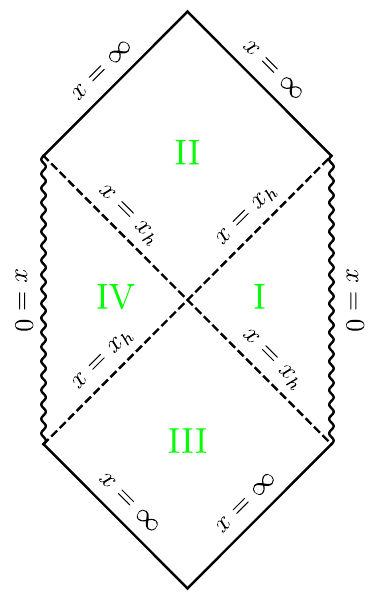}
		\caption{The Penrose diagram for $\Lambda=0$}
		\label{qBTZ2}
	\end{figure}

\begin{figure}[!htb]
		\includegraphics [width=0.25\textwidth]{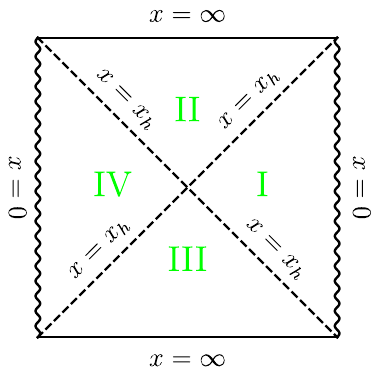}
		\caption{The Penrose diagram for $0 < \Lambda < 1/(2\zeta^2)$}
		\label{qBTZ3}
	\end{figure}

	{\it Physical properties---}We can calculate the ADM mass \cite{Arnowitt:1962hi, Carlip:1998uc} from Eqs. \eqref{SWCgauge} and \eqref{feff}, yielding $M = m(1-2\zeta^2\Lambda)$. 
	The parameter $m$ is considered positive to ensure compatibility with the classical case. 
	Consequently, to guarantee $M > 0$, the cosmological constant $\Lambda$ must be less than $1/(2\zeta^2)$. 
	Therefore, our analysis of the structure of spacetime \eqref{SWCgauge} is restricted to $\Lambda < 1/(2\zeta^2)$. 
	The equation $f_{\mathrm{eff}}(x) = 0$ can have up to two positive solutions: $x_1 = 2\sqrt{2Gm}/\sqrt{-\Lambda}$, $x_2 = 2\sqrt{2Gm}\zeta/\sqrt{1-\zeta^2 \Lambda}$. 
	The resulting geometry falls into two categories depending on the value of $\Lambda$. 
	\begin{enumerate}[label=(\roman*)]
		\item $\Lambda < 0$. In this scenario, the spacetime describes a black hole with two horizons, 
		an outer one at $x_+ = x_1$ and an inner one at $x_- = x_2$, 
		which reduces to the well-known BTZ black hole when $\zeta = 0$. 
		The classical singularity is resolved by LQG-inspired effects, 
		which induce a transition region connecting the black hole to a white hole. 
		This behavior is analogous to that appearing in some loop quantum black hole models \cite{Modesto:2008im, Kelly:2020lec, Munch:2021oqn, Lewandowski:2022zce}. 
		The Penrose diagram is shown in Fig. \ref{qBTZ1}. 
		\item $0 \leq \Lambda < 1/(2\zeta^2)$. 
		In this regime, although no black hole solution is allowed, a cosmological horizon emerges at $x_h = x_2$. 
		This is significantly different from the classical picture. 
		The Penrose diagram for $\Lambda = 0$ is depicted in Fig. \ref{qBTZ2}, 
		while that for $0 < \Lambda < 1/(2\zeta^2)$ is illustrated in Fig. \ref{qBTZ3}. 
		The primary difference is that the spacetime is asymptotically flat for $\Lambda = 0$, 
		while it becomes asymptotically de Sitter for $0 < \Lambda < 1/(2\zeta^2)$. 
	\end{enumerate}
	
	The calculation of the Ricci scalar yields 
	\begin{equation}
		R = 6\Lambda - 6\zeta^2 \Lambda^2 - \frac{128 G^2 m^2 \zeta^2}{x^4},
	\end{equation}
	which clearly shows that $x = 0$ is a singularity in both cases. 
	Nevertheless, this singularity can never be hit by timelike geodesics. 
	A massive particle moving along a timelike geodesic toward the singularity 
	will be reflected outward at some finite distance, 
	indicating that the spacetime is timelike geodesically complete. 
	Although non-geodesic timelike curves can extend to the singularity, 
	it remains inaccessible to any physical observer. 
	This is because any such path would demand an infinite integrated acceleration, 
	a requirement that is infeasible for a physically reasonable observer carrying a finite payload \cite{Chakrabarti:1983cf}.

	{\it Summary and discussions---}We have successfully constructed three independent LQG-inspired frameworks for studying (2+1)-dimensional black hole solutions: the $\bar{\mu}$-scheme holonomy corrections, the quantum Oppenheimer--Snyder collapse, and a covariant effective model. The three constructions presented in this work, while sharing the same holonomy-modified input motivated by LQC, are methodologically distinct. Their differences are most clearly seen through the lens of covariance, one of the central challenges for effective loop quantum black hole models. The $\bar{\mu}$ scheme fixes the space-time gauge and solves the radial diffeomorphism constraint before incorporating the holonomy modifications. As noted in the literature \cite{Bojowald:2020dkb, Belfaqih:2024dzn, Belfaqih:2026qgj}, the modified Hamiltonian constraint becomes the only remaining gauge generator, generating time translations but not radial diffeomorphisms; as a result, covariance cannot be systematically examined or ensured. The quantum Oppenheimer-Snyder model provides an independent construction. It starts from the dynamics of a collapsing dust ball governed by the LQC-effective Friedmann equation and matches the interior to a stationary exterior vacuum via the Darmois-Israel junction conditions. The junction condition being a tensor equation, the exterior geometry obtained from this matching is covariant \cite{Yang:2025ufs}: different coordinate choices in the exterior region yield the same metric tensor. This method is therefore distinct from the $\bar{\mu}$ scheme in both its physical setup and its technical approach, and it serves as a nontrivial independent cross-check. The covariant effective model, by contrast, is built from first principles by retaining the full set of gauge generators and requiring that the gauge transformations of the effective metric coincide with the spacetime diffeomorphism transformations. This ensures that covariance is systematically guaranteed. Remarkably, despite these substantial differences, all three approaches converge on the same qualitative causal structure, which strengthens the robustness of our conclusions.
    
    All three approaches yield LQG-inspired corrected spacetimes with the same qualitative causal structure, notably the emergence of a cosmological horizon in the asymptotically flat case ($\Lambda=0$). For $\Lambda<0$, they reproduce a effective-corrected BTZ black hole in which the classical singularity is resolved by a transition region connecting the black hole to a white hole. For $\Lambda>0$, they give rise to a cosmological horizon in an asymptotically de Sitter spacetime. In all three regimes, the spacetime is timelike geodesically complete \cite{Banados:1992gq}. Our results show that the LQG-inspired corrected (2+1)-dimensional solutions fill the gap that existed in the classical case by populating the previously empty sectors with non-trivial spacetime geometries. In this sense, the asymmetry between classical (2+1)- and (3+1)-dimensional solutions is eliminated when effective quantum gravity effects are taken into account.
    
    The emergence of a cosmological horizon at $\Lambda=0$ is especially noteworthy. In the classical theory, $\Lambda=0$ corresponds to a trivial vacuum without any horizon. While mechanisms such as the introduction of charge \cite{Cataldo:2000we}, can generate a cosmological horizon, our model offers a fundamentally distinct mechanism originating purely from effective quantum gravity effects. This horizon arises from the intrinsic repulsion of quantum geometry, which drastically modifies the causal structure of spacetime.
    
    Our work thus establishes that the key qualitative features of the LQG-inspired corrected BTZ geometry are robust across different loop-quantization prescriptions, and that the covariant effective framework consistently captures these features while preserving general covariance, paving the way for further investigations of lower-dimensional quantum gravity models. The present result of the covariant effective model is obtained under static configurations. A proper treatment of non-static cases would require further investigation, which we leave for future work.


	\begin{acknowledgments}
		We acknowledge the valuable discussions with Hongbao Zhang, Yongge Ma and Cong Zhang. 
		This work is supported by National Natural Science Foundation of China (NSFC) with Grants No.12275087.
	\end{acknowledgments}

	\bibliographystyle{unsrt}

	\section{Supplemental Material}\label{Supplement}

	\subsection{Derivation of the Hamiltonian formulation}\label{Hamiltonianformulation}
	A general circularly symmetric spacetime can be described by the metric \cite{Kelly:2020uwj}: 
	\begin{equation}
		\mathrm{d}s^2 = -N^2 \mathrm{d}t^2 + f^2 (\mathrm{d}x + N^x \mathrm{d}t)^2 + h^2 \mathrm{d}\phi^2 .
	\end{equation}
	The spatial metric can be expressed as  $q_{ab} = \delta_{ij} e_a^i e_b^j$, with the co-triads being: 
	\begin{equation}
		e_x^1 = f, \quad e_\phi^2 = h .
	\end{equation}
	The densitized triads are defined as $E_i^a = \sqrt{\det(q)} e_i^a$, and we denote the non-vanishing components as: 
	\begin{equation}
		E^1 \equiv E_1^x = h, \quad E^2 \equiv E_2^\phi = f .
	\end{equation}
	Then, the metric can be rewritten as: 
	\begin{equation}
		\mathrm{d}s^2 = -N^2 \mathrm{d}t^2 + (E^2)^2 (\mathrm{d}x + N^x \mathrm{d}t)^2 + (E^1)^2 \mathrm{d}\phi^2 .
	\end{equation}
	The extrinsic curvature is $K_{ab} = \frac{1}{2} \mathcal{L}_t q_{ab}$, 
	and the corresponding one-form is $K_a^i = K_{ab} e^{bi}$, with the non-vanishing components denoted as: 
	\begin{equation}
		K_1 \equiv K_x^1, \quad K_2 \equiv K_\phi^2 .
	\end{equation}

	The action for the vacuum gravitational field is the Einstein-Hilbert action \cite{Carlip:1998uc}: 
	\begin{equation}
		\mathcal{S} = \frac{1}{2\kappa} \int d^3X \sqrt{-\det(g)} (R - 2\Lambda) ,
	\end{equation}
	where $\kappa = 8 \pi G$. 
	From this action, general relativity can be reformulated into its Hamiltonian formulation, 
	as detailed in \cite{Wald:1984rg, Liang:2023ahd, Thiemann:2007pyv}. 
	Following the analysis in \cite{Thiemann:2007pyv}, we obtain: 
	\begin{equation}
		\mathcal{S} = \frac{1}{2\kappa} \int \mathrm{d}t \int \mathrm{d}^{2} X \left( 2 \dot{K}_a^i E_i^a - \left[ N^a H_a + N H \right] \right) ,
	\end{equation}
	where
	\begin{equation}
		H_a = -2D_b \left[ K_a^i E_i^b - \delta_a^b K_c^i E_i^c \right] ,
	\end{equation}
	\begin{equation}
		H = \frac{1}{\sqrt{\det(q)}} \left(K_a^j K_b^i - K_a^i K_b^j\right) E_i^a E_j^b - \sqrt{\det(q)}(R - 2\Lambda) ,
	\end{equation}
	and a dot over $K$ means derivative with respect to $t$. 
	Utilizing the coordinate choices we have just presented and integrating over $\mathrm{d}\phi$, 
	we arrive at the symmetry-reduced action: 
	\begin{equation}
		\mathcal{S} = \int \mathrm{d}t \int \mathrm{d}x \left( \frac{E^1 \partial_t K_1 + E^2 \partial_t K_2}{4G} - \left[ N^x H_x + N H \right] \right) ,
	\end{equation}
	where
	\begin{equation}
		H_x = \frac{1}{4G}\left(E^2 \partial_x K_2 - K_1 \partial_x E^1\right),
	\end{equation}
	and
	\begin{equation}
		H = -\frac{K_1 K_2}{4G} 
		- \frac{(\partial_x E^1) \partial_x E^2}{4G(E^2)^2} 
		+ \frac{\partial_x^2 E^1}{4GE^2} 
		+ \frac{E^1 E^2 \Lambda}{4G}.
	\end{equation}

	\subsection{Effective dynamics in $\bar{\mu}$ scheme}\label{Effectivedynamics}
	Based on the Hamiltonian formulation in \ref{Hamiltonianformulation}, we adopt a partial gauge-fixing by setting $E^1 = x$. 
	The constraint $\chi \equiv E^1 - x = 0$ is second-class with respect to $H_x$, 
	and thus we can use this condition to gauge fix $H_x$, yielding 
	\begin{equation}
		E^1 = x, \qquad K_1 = E^{2}\partial_x K_{2} .
	\end{equation}
	In order for the gauge to be preserved dynamically, $\dot\chi = 0$, one must have $\dot{E^1}=0$ and therefore 
	\begin{equation}
		N^{x} = -N K_2.
	\end{equation}
	With this gauge, the action simplifies considerably to: 
	\begin{equation}
		\mathcal{S}_{GF} = \int \mathrm{d}t \int \mathrm{d}x \left( \frac{E^2 \partial_t K_2}{4G} - N H \right) ,
	\end{equation}
	with
	\begin{equation}
		H = -\frac{K_2 E^2 \partial_x K_2}{4G} 
		- \frac{\partial_x E^2}{4G(E^2)^2} 
		+ \frac{x E^2 \Lambda}{4G}.
	\end{equation}
    The remaining fundamental Poisson bracket is:
    \begin{equation}\label{fPB_GF}
    \{K_2(x), E^2(y)\} = 4G\delta(x, y),
    \end{equation}
    and the constraint algebra also simplifies accordingly:
    \begin{align}\label{HH}
    \{H[N_1], H[N_2]\}
    &= H[-(N_1\partial_x N_2 - N_2\partial_x N_1)K_2] \notag \\
    &= H[ N^{x}_1\partial_x N_2 - N^{x}_2\partial_x N_1 ].
    \end{align}
    Then, the equations of motion are
    \begin{equation}
    \dot{E}^{2} = -K_2 \partial_x(N E^{2}),
    \end{equation}
    \begin{equation}
    \dot{K_2} = \frac{\partial_x N}{(E^{2})^{2}} - N K_2 \partial_x K_2 + N x \Lambda.
    \end{equation}

    Incorporating holonomy corrections in the $\bar{\mu}$ scheme, we replace the connection $K_2$ by its holonomy:
    \begin{equation}
    K_2 \to \frac{x}{\zeta}\, \sin\left( \frac{\zeta}{x}K_2 \right),
    \end{equation}
    where $\zeta=\gamma L$ is the quantum parameter.
    With this replacement, the effective Hamiltonian constraint reads 
    \begin{equation}\label{HCLQG}
    H^{(LQG)} = -\frac{1}{4G} \bigg[\frac{E^{2}}{2}\partial_x\left(\frac{x^{2}}{\zeta^2}\sin^{2}\frac{\zeta K_2}{x} \right) + \frac{\partial_x E^{2}}{(E^{2})^{2}} - x E^2 \Lambda \bigg].
    \end{equation}
    The constraint algebra obtained from the Poisson bracket of the effective Hamiltonian constraint with itself takes the form:
    \begin{equation}\label{HLQGHLQG}
		\begin{aligned}
			& \quad \{H^{(LQG)}[N_1],H^{(LQG)}[N_2]\} \\
			& = H^{(LQG)}\left[ -\frac{x}{\zeta}\sin\frac{\zeta K_2}{x}\cos\frac{\zeta K_2}{x}(N_1\partial_x N_2 - N_2\partial_x N_1) \right].
		\end{aligned}
	\end{equation}
    Comparing the classical constraint algebra \eqref{HH} with its effective counterpart \eqref{HLQGHLQG} suggests taking
    \begin{equation}\label{SVLQG}
    N^{x} = -\frac{N x}{\zeta}\sin\frac{\zeta K_2}{x}\cos\frac{\zeta K_2}{x}
    \end{equation}
    to ensure the effective constraint algebra retains the classical form
    \begin{equation}
		\begin{aligned}
			& \quad \{H^{(LQG)}[N_1], H^{(LQG)}[N_2]\} \\
			& = H^{(LQG)}[ N^{x}_1\partial_x N_2 - N^{x}_2\partial_x N_1 ].
		\end{aligned}
	\end{equation}
    With the shift vector \eqref{SVLQG}, the effective metric takes the form
    \begin{equation}
		\mathrm{d}s^2 = -N^2 \mathrm{d}t^2 + (E^2)^2 (\mathrm{d}x + N^x \mathrm{d}t)^2 + x^2 \mathrm{d}\phi^2.
	\end{equation}
    The equations of motion for $E^2$ and $K_2$ follow directly from the effective Hamiltonian constraint \eqref{HCLQG} and the Poisson bracket \eqref{fPB_GF}:
    \begin{equation}
    \dot{E}^{2} = -\frac{x}{\zeta}\partial_x\left(N E^{2}\right)\sin\frac{\zeta K_2}{x}\cos\frac{\zeta K_2}{x},
    \end{equation}
    \begin{equation}
    \dot{K_2} = \frac{\partial_x N}{(E^{2})^{2}} - \frac{N}{2\zeta^2}\partial_x\left( x^{2} \sin^{2}\frac{\zeta K_2}{x} \right) + N x \Lambda.
    \end{equation}

    We look for a stationary solution of the equations of motion and the effective Hamiltonian constraint $H^{(LQG)} = 0$ in Painlev\'e--Gullstrand-like coordinates with lapse $N=1$.
    From $\dot E^2 = 0$, we can set
    \begin{equation}
    E^{2} = 1,
    \end{equation}
    and from $\dot K_2 = 0$,
    \begin{equation}
    K_2 = -\frac{x}{\zeta}\arcsin\sqrt{\frac{C}{x^{2}}+\zeta^2\Lambda},
    \end{equation}
    where $C$ is a constant of integration. The constant can be fixed as $C=\zeta^2(1+8Gm)$ by considering the solution at large $x$ where the classical BTZ solution should be recovered. Finally, the shift vector is determined as 
    \begin{equation}
    N^{x} = \sqrt{ 1 + 8Gm + \Lambda x^{2} - \frac{\zeta^2}{x^{2}} (1 + 8Gm + \Lambda x^{2})^2 },
    \end{equation}
    and the effective metric is 
    \begin{equation}
		\begin{aligned}
			& \mathrm{d}s^{2} = \\
			& -\left( -8Gm - \Lambda x^{2} + \frac{\zeta^2}{x^{2}} (1+8Gm + \Lambda x^{2})^2 \right)\mathrm{d}T^{2} \\
            & +2\sqrt{ 1 + 8Gm + \Lambda x^{2} - \frac{\zeta^2}{x^{2}} (1 + 8Gm + \Lambda x^{2})^2 }\mathrm{d}T\mathrm{d}x \\
            & +\mathrm{d}x^{2} + x^{2} \mathrm{d}\phi^{2} .
		\end{aligned}
	\end{equation}

	\subsection{(2+1)-dimensional quantum Oppenheimer-Snyder model}\label{Collapse}
	We consider a disk of pressureless, collapsing dust surrounded by a vacuum region. 
	The interior spacetime of the dust can be described by the spatially flat FRW metric: 
	\begin{equation}\label{dsin}
		\mathrm{d}s_{in}^2=-\mathrm{d}\tau^2+a^2(\tau)[\mathrm{d}\chi^2+\chi^2\mathrm{d}\phi^2] .
	\end{equation}
	The induced metric on the dust surface $\mathscr{S}$ is: 
	\begin{equation}
		\left.\mathrm{d}s_{in}^2\right|_\mathscr{S}=-\mathrm{d}\tau^2+a^2(\tau)\chi_0^2\mathrm{d}\phi^2 .
	\end{equation}
	The exterior spacetime is given by: 
	\begin{equation}
		\mathrm{d}s_{out}^2=-f(x)\mathrm{d}t^2+h(x)^{-1}\mathrm{d}x^2+x^2\mathrm{d}\phi^2 ,
	\end{equation}
	where we can also derive the metric on the dust surface as: 
	\begin{equation}
		\left.\mathrm{d}s_{out}^2\right|_\mathscr{S}=-(f\dot{t}^2-h^{-1}\dot{x}^2)\mathrm{d}\tau^2+x^2(\tau)\mathrm{d}\phi^2 .
	\end{equation}
	Here, $t$ and $x$ are both $\tau$-dependent, and $\dot{t}$ and $\dot{x}$ are derivatives with respect to $\tau$. 
	According to the Darmois-Israel junction conditions, 
	the first and second fundamental forms of the induced metric obtained from the interior and exterior must be equal. 
	This yields: 
	\begin{equation}\label{g00}
		1=f\dot{t}^2-h^{-1}\dot{x}^2 ,
	\end{equation}
	\begin{equation}\label{g11}
		a(\tau)\chi_0=x(\tau) ,
	\end{equation}
	and 
	\begin{equation}\label{K11}
		a(\tau)\chi_0 = K_{\phi\phi} = xE\sqrt{f^{-1}h} ,
	\end{equation}
	where $E = -\left(\frac{\partial}{\partial t}\right)_a\left(\frac{\partial}{\partial \tau}\right)^a$ is the conserved energy and $K_{\phi\phi}$ is the extrinsic curvature. 
	A straightforward calculation then gives: 
	\begin{equation}\label{fg}
		f = h = 1-\mathcal{H}^2x^2 ,
	\end{equation}
	where $\mathcal{H}=\dot{a}/a$ is the Hubble parameter, and $E=1$ is chosen. 
	The loop quantum cosmology in 2+1 dimensions has already been established in \cite{Zhang:2014xqa}. 
	The resulting effective Friedmann equation is: 
	\begin{equation}\label{QFE}
		\mathcal{H}^2=\kappa\left(\rho+\frac{\Lambda}{\kappa}\right)\left(1-\frac{\rho+\frac{\Lambda}{\kappa}}{\rho_c}\right) ,
	\end{equation}
	where $\rho$ is the matter density, and 
	\begin{equation}\label{rhoc}
		\begin{aligned}\rho_c&=\frac{1}{\kappa\gamma^2L^2}\end{aligned}
	\end{equation}
	is the critical matter density, with $\gamma$ and $L$ being the Barbero-Immirzi parameter and the minimal length gap, respectively. 
	Substituting Eq. \eqref{QFE} into Eq. \eqref{fg} yields: 
	\begin{equation}\label{fout}
    \begin{split}
        f(x)=h(x)=1+(-8Gm_d - \Lambda x^{2}) \\
        \times \left[1 + \frac{\zeta^2}{x^{2}} (-8Gm_d - \Lambda x^{2})\right] ,
    \end{split}
	\end{equation}
	where $m_d = \pi \rho a^2 \chi_0^2$ is the total mass of the dust and we have used the relation $\zeta = \gamma L$. If we wish to recover the standard form of the BTZ solution at $\zeta=0$, we simply need to define the mass parameter \cite{Ross:1992ba} $m=m_d-\frac{1}{8 G}$. Then, the metric functions become:
    \begin{equation}
		f(x)=h(x)=(-8Gm - \Lambda x^{2}) + \frac{\zeta^2}{x^{2}} (1+8Gm + \Lambda x^{2})^2 .
	\end{equation}

	\subsection{Necessary and sufficient conditions for covariance in effective quantum case}\label{NSconditions}
	We begin with the equation of motion in the form of a Lie derivative: 
	\begin{equation}\label{LieEOM}
		\mathcal{L}_{\mathfrak{N}} \mathscr{F} = \{\mathscr{F}, H_{\mathrm{eff}}[N]\} ,
	\end{equation}
	whose solutions define curves $\mathscr{F}^{(t)}(x)$ in phase space. 
	Next, we consider the infinitesimal gauge transformation for 
	$g_{\rho\sigma}^{(\mu)}$ generated by $H[\alpha N] + H_x[\beta^x]$. 
	This transformation maps a phase-space curve satisfying \eqref{LieEOM} to a new one: 
	\begin{equation}
		\mathscr{F}^{(t)}(x) \to \mathscr{F}^{(t)}(x) + \epsilon \delta \mathscr{F}^{(t)}(x) + O(\epsilon^2), 
	\end{equation}
	where $\delta \mathscr{F} := \{ \mathscr{F}, H_{\mathrm{eff}}[\alpha N] + H_x[\beta^x] \}$. 
	According to \cite{Zhang:2024khj, Zhang:2024ney}, it is sufficient to consider the case where
	$\alpha$, $\beta^x$, $N$, and $N^x$ are all phase-space-independent. 
	The results also hold on-shell if these quantities depend on phase space. 
	By definition, we have:  
	\begin{equation}\label{deltaE1}
		\begin{aligned}
			\delta E^1 &:= \{ E^1, H_{\mathrm{eff}}[\alpha N] + H_x[\beta^x] \} \\
			&= \mathcal{L}_{\alpha \mathfrak{N} + \beta} E^1 - \Delta_1,
		\end{aligned}
	\end{equation}
	\begin{equation}\label{deltaE2}
		\begin{aligned}
			\delta E^2 &:= \{ E^2, H_{\mathrm{eff}}[\alpha N] + H_x[\beta^x] \} \\
			&= \mathcal{L}_{\alpha \mathfrak{N} + \beta} E^2 - E^2 \mathfrak{N}^\rho \partial_\rho \alpha - \Delta_2,
		\end{aligned}
	\end{equation}
	where
	\begin{equation}\label{DeltaI}
		\Delta_I = \sum_{n \geq 1} (-1)^n \left[ \sum_{m=1}^n \binom{n}{m} (\partial_x^m \alpha) \frac{\partial^n \left( N(x) \frac{\partial H_{\text{eff}}(x)}{\partial (\partial_x^n K_I(x))} \right)}{\partial x^{n-m}} \right] .
	\end{equation}
	The new curve in phase space satisfies equation of motion \eqref{LieEOM} 
	with respect to a new lapse function $N + \epsilon \delta N$ and a new shift vector $(N^x + \epsilon \delta N^x)\partial_x$. 
	The variations $\delta N$ and $\delta N^x$ are given by: 
	\begin{equation}\label{deltaNx}
		\delta N^x = - N^2 \mu S \partial_x \alpha - (\mathcal{L}_\beta \mathfrak{N})^x,
	\end{equation}
	\begin{equation}\label{deltaN}
		\delta N = \mathcal{L}_{\alpha \mathfrak{N} + \beta} N + N \mathfrak{N}^\rho \partial_\rho \alpha.
	\end{equation}
	Making use of Eqs. \eqref{deltaE1}, \eqref{deltaE2}, \eqref{deltaNx}, and \eqref{deltaN}, we obtain: 
	\begin{equation}\label{qcovar}
		\begin{aligned}
			&\delta g_{\rho\sigma}^{(\mu)} \mathrm{d}x^{\rho}\mathrm{d}x^{\sigma} \\
			&= \mathcal{L}_{\alpha\mathfrak{N}+\beta} (g_{\rho\sigma}^{(\mu)} \mathrm{d}x^{\rho}\mathrm{d}x^{\sigma}) - 2 E^1 \Delta_1 \mathrm{d}\phi^2 \\
			&\quad + \frac{(E^2)^{2}}{\mu^2} \left(\mathcal{L}_{\alpha\mathfrak{N}+\beta} \mu - \delta\mu - \frac{2\mu\Delta_2}{E^2}\right) (\mathrm{d}x + N^x \mathrm{d}t)^2.
		\end{aligned}
	\end{equation}
	To ensure covariance, 
	the last two terms in Eq. \eqref{qcovar} must vanish, leading to: 
	\begin{equation}\label{Delta1}
		\Delta_1 = 0 ,
	\end{equation}
	\begin{equation}\label{Delta2}
		\mathcal{L}_{\alpha\mathfrak{N}+\beta} \mu - \delta\mu - \frac{2\mu\Delta_2}{E^2} = 0 ,
	\end{equation}
	where $\delta \mu = \{ \mu, H_{\mathrm{eff}}[\alpha N] + H_x[\beta^x] \}$.
	Combining Eqs. \eqref{DeltaI} and \eqref{Delta1}, we get the first condition: 
	$H_{\mathrm{eff}}$ is independent of $\partial_x^n K_1$ for all $n \geq 1$. 
	Combining Eqs. \eqref{deltaE2} and \eqref{Delta2}, we get the second condition: 
	The equation $\alpha\{\mu S, H_{\mathrm{eff}}[N]\} = \{\mu S, H_{\mathrm{eff}}[\alpha N]\}$
	holds for all phase-space-independent $\alpha$ and $N$.

    \subsection{Black hole thermodynamics}\label{Thermodynamics}
	If we consider the thermodynamic properties based on Eqs. \eqref{SWCgauge} and \eqref{feff}, 
	we will find they are the same as those of the classical case. 
	It is because black hole thermodynamics is a theory related to the horizon \cite{Kubiznak:2016qmn}.
	The outer horizon $x_+$ is unaffected by effective corrections, so are the thermodynamic properties. 
	However, it should be noted that the parameter $m$ is not yet justified as the ADM mass. 
	Therefore, we can recast the metric function in terms of ADM mass as follows: 
	\begin{equation}
		f_{\mathrm{eff, M}}(x) = -8 G M - \Lambda x^2 (1 - \zeta^2 \Lambda) + \frac{64 G^2 M^2 \zeta^2}{x^2 (1 - 2 \zeta^2 \Lambda)^2}.
	\end{equation}
	In this formulation, the black hole's outer horizon is denoted by $x_{o}$. 
	Using the the units $G = \hbar = c = k_B = 1$, the generalized first law \cite{Kubiznak:2016qmn} is modified to: 
	\begin{equation}\label{MFL}
		\delta M = (1 - 2 \zeta^2 \Lambda) \frac{\varkappa}{8\pi} \delta A + (1 - 4 \zeta^2 \Lambda) V \delta P,
	\end{equation}
	where $\varkappa = f_{\mathrm{eff, M}}'(x_{o}) / 2$ is the surface gravity, $A = 2 \pi x_{o}$ is the horizon area, 
	$V = \pi x_{o}^2$ is the thermodynamic volume, and $P = -\Lambda / (8\pi)$ is the pressure. 
	The presence of the $\zeta$-dependent factors in Eq. \eqref{MFL} represents 
	the incorporation of LQG effects into black hole thermodynamics.

\end{CJK}
\end{document}